\documentclass[11pt,twoside]{article}

\usepackage{asp2014}

\aspSuppressVolSlug
\resetcounters

\begin{document}

\title{TOPCAT/STILTS Integration}

\author{Mark~Taylor}
\affil{H.~H.~Wills Physics Laboratory, Tyndall Avenue,
       University of Bristol, UK;
       \email{m.b.taylor@bristol.ac.uk}}

\paperauthor{Mark~Taylor}{m.b.taylor@bristol.ac.uk}{0000-0002-4209-1479}{University of Bristol}{School of Physics}{Bristol}{Bristol}{BS8 1TL}{U.K.}



\begin{abstract}
TOPCAT and STILTS are related packages for desktop analysis of
tabular data, presenting GUI and command-line interfaces
respectively to much of the same functionality.
This paper presents features in TOPCAT that facilitate use of STILTS.
\end{abstract}



\section{Introduction}

TOPCAT\footnote{\url{http://www.starlink.ac.uk/topcat/}}
\citep{2005ASPC..347...29T}
is an established interactive desktop GUI tool for data analysis
of tables, offering visualisation, crossmatching, data manipulation,
and access to VO services among other capabilities.
STILTS\footnote{\url{http://www.starlink.ac.uk/stilts/}}
\citep{2006ASPC..351..666T}
is a suite of command-line tools
providing a scriptable interface to many of the same capabilities,
built on the same software infrastructure.
While scripting is a powerful way to approach many data analysis tasks,
the learning curve for STILTS is rather steeper than for its point'n'click
counterpart, with the result that some TOPCAT users may be reluctant
to exploit the scripting capabilities on offer because of the
difficulty of learning to use them.

To ease the transition from GUI to command-line operation,
it is therefore desirable for TOPCAT to offer some way of
assisting users to invoke STILTS.
There are various ways that this could be imagined;
one model would be for TOPCAT to log all the GUI operations
performed by the user as a sequence of STILTS commands
that could be replayed to achieve the same end effect.
Unfortunately, although equivalent operations exist in many cases
for the two UIs, the workflows in the two environments are too
dissimilar to make this work well from both an implementation
and a usability point of view.

Instead the approach taken is for various TOPCAT windows to report
the STILTS command corresponding to the operation currently
configured.
Such commands can then be copied and pasted onto the command line
or into a script so that the current TOPCAT operation can be
performed programmatically.
This can be used either directly to duplicate GUI behaviour,
or as a template for users to adapt when writing their own
scripts performing similar operations on the same or other data.

In this way users can build a STILTS command line
by interacting with selection boxes and other
relatively friendly graphical elements
rather than having to supply a list of parameter key-value pairs
from scratch.

\articlefigure[scale=0.52]
              {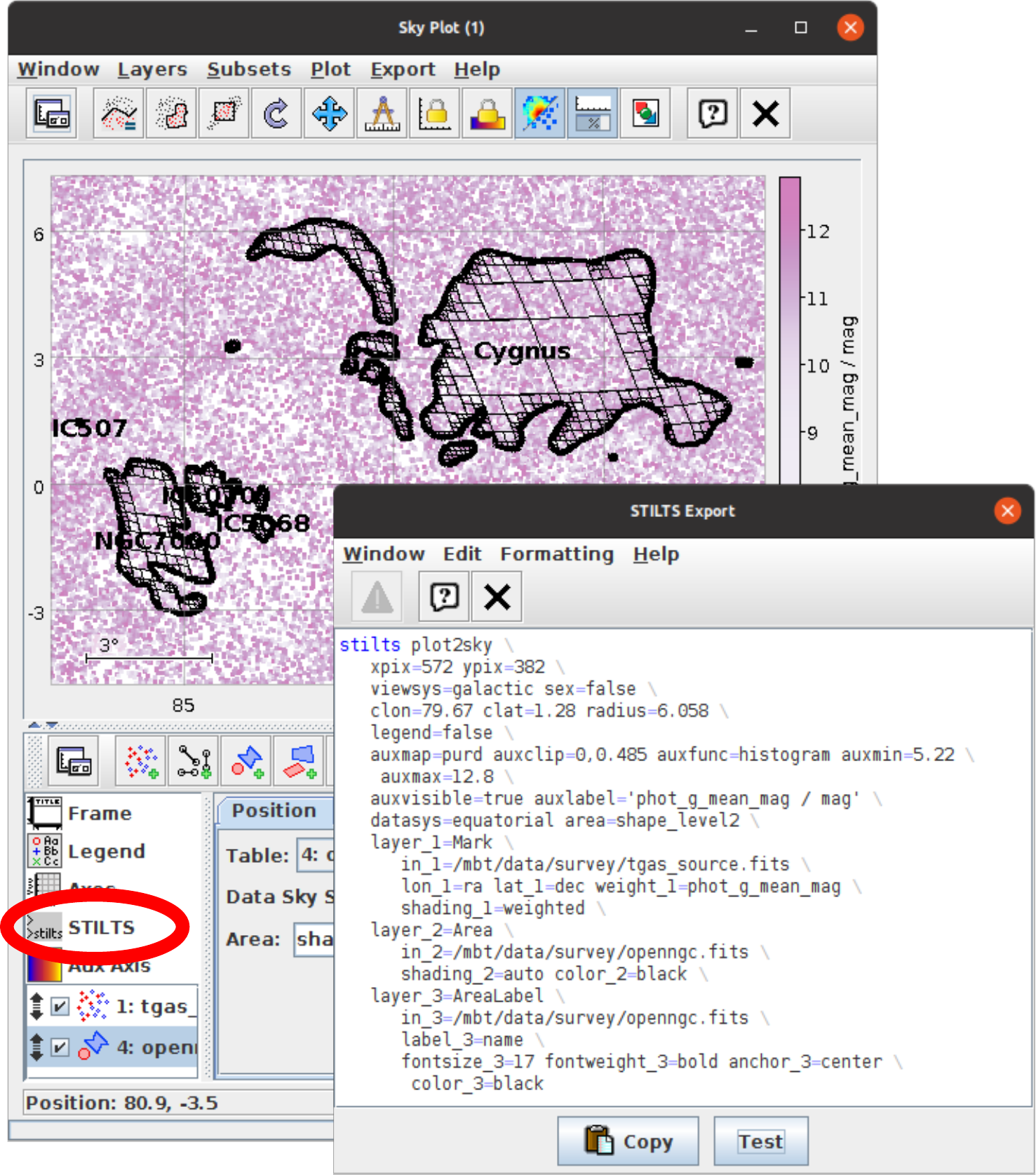}
              {P409-plotfig}
              {Plot window showing STILTS equivalent command.
               Selecting the highlighted control displays the command
               either in the window or in a popup as shown.
               Running the displayed command from the shell
               will reproduce the visible plot.}

\articlefigure[scale=0.52]
              {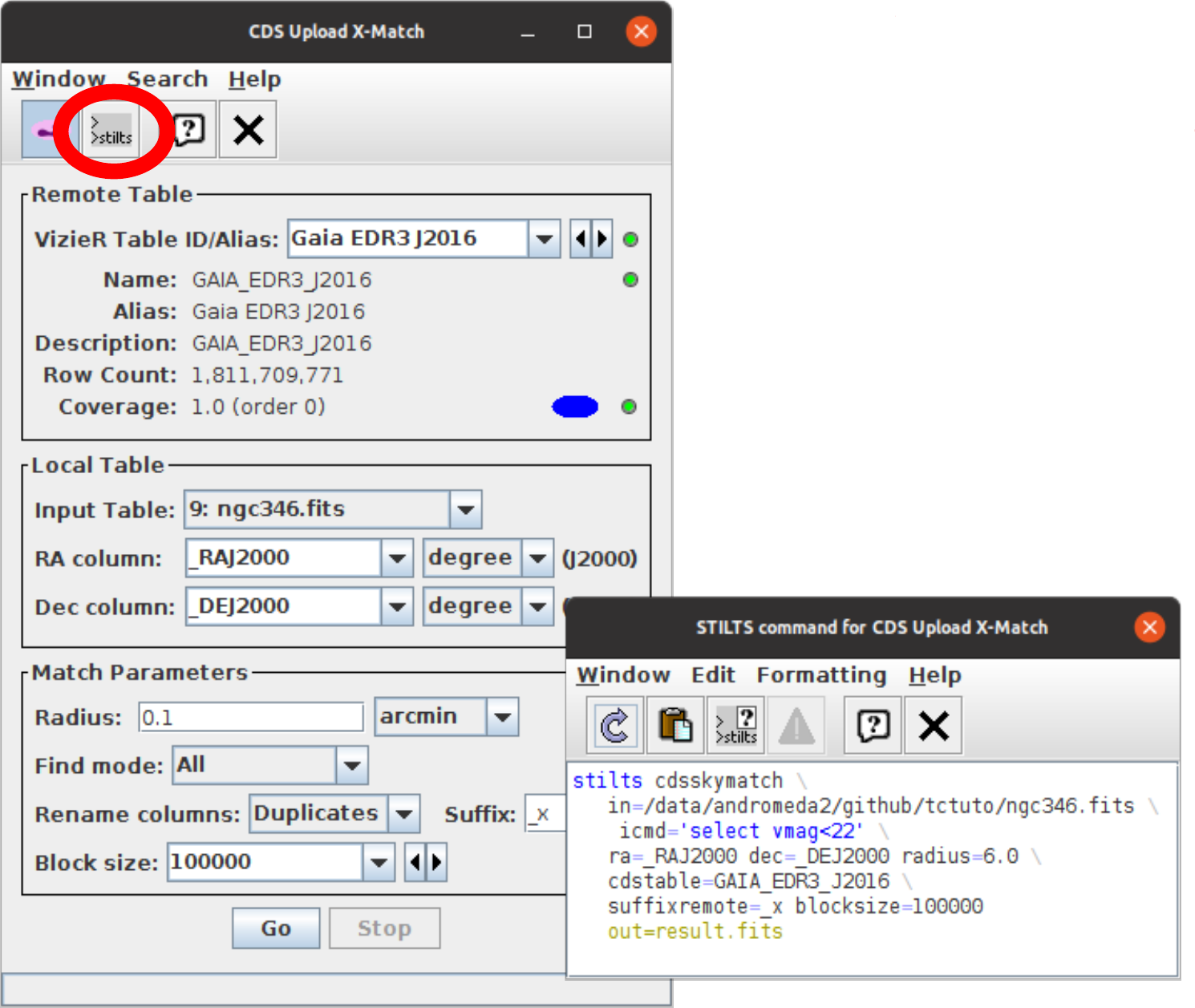}
              {P409-matchfig}
              {CDS X-Match window showing STILTS equivalent command.
               When the highlighted toolbar button is clicked
               the popup window appears.
               Running the displayed command from the shell
               will perform the same remote crossmatch
               operation that the {\em Go\/} button executes.}

\section{Implementation Status}

The visualisation windows in TOPCAT have offered the
STILTS command reporting feature since v4.5 (2017),
see Figure~\ref{P409-plotfig}.
Visualisation was in fact the most pressing context for this behaviour
since the STILTS plotting commands can be quite complicated,
so that application help in constructing them is especially welcome.
Since the recent release v4.10-2 (2024)
this feature has been extended to a number of other windows
as well, covering crossmatches of various kinds and interaction
with external (mostly Virtual Observatory) services,
see Figure~\ref{P409-matchfig}.

Specifically, the TOPCAT windows involved and corresponding STILTS
commands are:
\begin{itemize}
\item All visualisation windows
      (commands {\tt plot2plane}, {\tt plot2sky}, {\tt plot2cube},
                {\tt plot2sphere}, {\tt plot2corner}, {\tt plot2time})
\item Pair, Internal and Multi-table Crossmatch windows
      (commands {\tt tmatch2}. {\tt tmatch1}, {\tt tmatchn}, {\tt tskymatch})
\item TAP window
      (command {\tt tapquery})
\item CDS Upload X-Match window
      (command {\tt cdsskymatch})
\item Single Cone Search, SIA, SSA windows
      (command {\tt tcone})
\item Multiple Cone Search, SIA, SSA windows
      (command {\tt coneskymatch})
\end{itemize}

To view the STILTS command corresponding to the current state
of a window the user presses a {\em STILTS\/} button in the toolbar,
and a dialogue pops up displaying the
command which should reproduce the action taken by the current
configuration.  The content of this dialogue is updated continuously
according to user interaction with the parent window.
The procedure is slightly different in the visualisation windows
for historical and ergonomic reasons;
clicking the STILTS control in the lower panel
displays the command within the parent window,
and another action can pop it out.

\section{Limitations}

The STILTS commands are reported on a best-efforts basis.
The presented text is a reconstruction by TOPCAT
of the most closely corresponding STILTS command,
it is not a simple serialization of the action taken by TOPCAT;
unfortunately the UIs differ too much to make that possible.
In most cases pasting the reported command into the shell works correctly,
but this is not guaranteed.
Even if the reported command does not execute without error however,
it provides a good starting point from which a working invocation
can be written.

One thing that can go wrong is
reference to state that cannot be represented on the command line;
for instance
row selections defined by expressions and tables that exist as files
can appear as STILTS parameter values,
but hand-drawn row selections cannot,
and dynamically-created tables may present difficulties.
To mitigate this, parts of the command line for which failure is
possible or likely are flagged in blue or red text respectively.
Dummy execution is also performed, and identifiable syntax errors
are flagged and can be inspected by use of an {\em Error\/} button.

Details of the command formatting also present a problem,
since the application is unaware of syntax details
such as line continuation characters that depend on the
shell in which the command will be executed.
The user is therefore offered various formatting options that
configure style for line endings, value quoting, indentation etc
in the displayed text.

It should be noted that there are many capabilities of STILTS
that are not covered by this functionality,
for instance most aspects of the highly flexible pipeline processing
that it offers.

\section{Summary}

Many windows in TOPCAT are now able to report
the corresponding STILTS command invocation,
which should aid users of the GUI tool to
make use also of the command-line package where appropriate.

It is hoped that this will enable more TOPCAT users to benefit from
features of STILTS such as scriptability, repeatability
and enhanced scalability.

\bibliography{P409}  


\end{document}